# Observation of Silicon Nitride Nanomechanical Resonator Actuation Using Capacitive Substrate Excitation

Gengyang Mu[1,2], Nikaya Snell[1,2], Chang Zhang[1,2], Xitong Xie[1,2], Radin Tahvildari[3,4], Arnaud Weck[1,2,3], Michel Godin[3,4], Raphael St-Gelais[1,2,3*]

[1]*Department of Mechanical Engineering, University of Ottawa, Ottawa, ON, Canada*
[2]*Centre for Research in Photonics, University of Ottawa, Ottawa, ON, Canada*
[3]*Department of Physics, University of Ottawa, Ottawa, ON, Canada*
[4]*Ottawa-Carleton Institute for Biomedical Engineering, Ottawa, ON, Canada*

**Abstract**: We observe the actuation of silicon nitride (SiN) nanomechanical resonators by electrical excitation of metal-dielectric-semiconductor (MDS) capacitors on their supporting silicon substrate. We develop first-principle models explaining this actuation mechanism by acoustic waves resulting from voltage-dependent electrostatic forces in the MDS capacitors. Models are developed for actuation in the charge accumulation (Ni-pSi) and charge depletion (Al-pSi) regimes. Experimental observations confirm our prediction that charge accumulation (Ni-pSi) is more efficient at actuation than charge depletion. For a 2 V actuation signal, Ni-pSi capacitors achieve 10 nm actuation amplitude in square (1.7 × 1.7 mm) low-stress (∼100 MPa) SiN membrane resonators. In this case, electrical power dissipation in the chip is on the order of 0.1 µW, and spurious heating is less than 1 mK. Both these values could be further reduced by doping the substrate to minimize resistive dissipation. The actuation method is remarkably simple and only requires attachment of wires to the chip with vacuum-compatible nickel paste, with no extra photolithography step. All the chips presented in this work are fabricated in-house, and a detailed fabrication procedure is provided.

## I. INTRODUCTION

Free-standing silicon nitride (SiN) membranes are widely used as Nano-Electromechanical System (NEMS) for their outstandingly low mechanical dissipation and high quality (Q) factor [1–9]; however, actuation remains challenging in certain situations. While external piezoelectric actuators are acceptable in laboratory settings, practical applications would benefit from more integrated actuation. Previously demonstrated methods for free-standing membranes actuation include electrostatic [10], magnetic [11] and optical [12]. Electrostatic and magnetic methods rely on voltage-induced forces and are suitable for various SiN membrane resonator geometries. However, the electrostatic method requires electrodes deposited on the SiN membrane which can lead to Q-factor reduction and often requires a non-integrated external electrode [13,14]. The magnetic method relies on external magnets [11,15], which can be bulky and hard to implement in practical applications. Optical actuation is efficient at actuating and simultaneously detecting displacements [12,16]. However, the forces are modest and are mostly applicable to low mass objects. Requirement of a stable high performance laser can also be a drawback in many practical situations.

Here we experimentally observe actuation of SiN drum resonators by exciting an alternative (AC) current in the underlying silicon frame through a metal-SiN-silicon double capacitor assembly shown in Fig. 1 (a, b). We hypothesize that such AC electrical current is able to actuate the membrane by creating acoustic waves in the substrate, which can originate from either (1) electrostatic attraction forces in the capacitors or (2) thermal expansion forces in the substrate. In the following, we present our experimental observations as well as first-principle models for these two possible mechanisms. In the electrostatic model, voltage-dependent mechanical stress arising from charge attraction in the Metal-SiN-Si capacitor structure creates acoustic waves that can travel in the substrate and actuate the membrane. Such effect was previously observed using p-n [17] and Schottky junctions [18] under reverse bias, but not in the context of a capacitively coupled metal-dielectric-semiconductor (MDS) assembly. In the thermal expansion model, periodic Joule heating creates a similar acoustic wave in the substrate. Similar effect has been analyzed previously in the context of electrostriction studies in silicon [19]. Our calculations show that that electrostatic attraction in the capacitors is most likely the dominant effect in the present configuration as it likely dominated over thermal effects by several orders of magnitude.

*raphael.stgelais@uottawa.ca

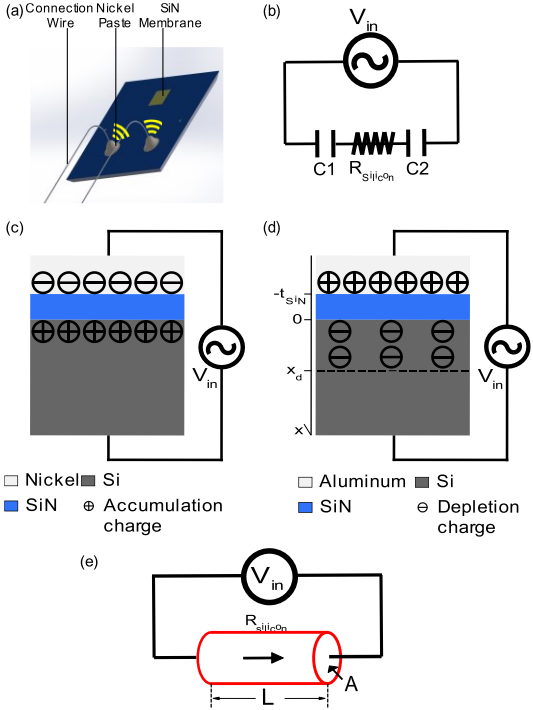

Figure 1. (a) SiN resonator actuation device schematic and (b) equivalent circuit. (c, d) Schematics of the capacitor charge distribution in the accumulation (c) regime and depletion (d) regime. (e) Equivalent simplified geometry and circuit for thermal expansion calculations.

## II. ELECTROSTATIC MODEL

In an MDS capacitor (see Fig. 1 c, d), varying the applied voltage influences the electrical charge distribution, which in turn can create varying mechanical stress resulting from electrostatic attraction forces. As within any solid object, we can expect that time-dependent strain will create an acoustic wave, which can travel in the substrate and reach the membrane, thus enabling actuation.

We analyze charge interaction inside MDS capacitors within the framework of two possible regimes—charge accumulation or charge depletion—which occur depending on the workfunction difference between the metal ($\psi_m$) and the semiconductor ($\psi_s$). Theoretical description of these regimes is widely available in the context of Metal-Oxide-Semiconductor (MOS) devices [20]. Unless otherwise noted, all descriptions below are for a p-type semiconductor substrate.

We first investigate stress in the accumulation regime (i.e., $\psi_m > \psi_s$, see Fig. 1c). In this case, charges accumulate in a sheet-like matter on each side of the dielectric (see Fig. 1c), and a simple parallel plate capacitor model suffices for calculating the mechanical stress. We define the capacitance per unit area as $C_{SiN} = \frac{\varepsilon_{SiN}}{t_{SiN}}$, with $\varepsilon_{SiN}$ as the dielectric constant for SiN, and $t_{SiN}$ as the dielectric thickness. The electrical energy stored in a capacitor, per unit area and for electrical potential $V$, is then given by $\frac{C_{SiN}V^2}{2}$. By taking the derivative of the capacitor energy along the thickness, we obtain the attractive force between the sheet charges, $F = \frac{C_{SiN}V^2 A}{2t_{SiN}}$, where $A$ is the metal contact area. We express the effective spring constant of the dielectric layer, in a one-dimensional approximation as $k = \frac{E_{SiN}A}{t_{SiN}}$, where $E_{SiN}$ denotes the Young's Modulus. By combing the expressions of force and stiffness, we obtain the strain energy stored in a capacitor in the accumulation regime ($U_{acc}$) as a function of voltage, in quasi-static conditions:

$$U_{acc} = \frac{F^2}{k} = \frac{V^4 \varepsilon_{SiN}^2 A}{4 t_{SiN}^3 E_{SiN}}. \quad (1)$$

The potential $V$ in Eq. (1) includes the applied potential, $V_{in}$, as well as the flat band potential $V_{fb} = \psi_m - \psi_s$, i.e., $V = V_{in} - V_{fb}$ [20]. Substituting in Eq. (1), we finally obtain:

$$U_{acc}(V_{in}) = \frac{(V_{in} - V_{fb})^4}{4} \frac{\varepsilon_{SiN}^2 A}{t_{SiN}^3 E_{SiN}}. \quad (2)$$

When applying a time-dependant actuation voltage $V_{in}$ on a single capacitor, the peak-to-peak variation in stored strain energy ($U_{p-p,acc}$) is given by:

$$\begin{aligned} U_{p-p,acc}(V_{in}) &= U_{acc}(-V_{in}) - U_{acc}(V_{in}) \\ &= \frac{2\varepsilon_{SiN}^2 A}{t_{SiN}^3 E_{SiN}} (V_{fb} V_{in}^3 + V_{fb}^3 V_{in}). \end{aligned} \quad (3)$$

In the equivalent circuit shown in Fig.1 (b), two capacitors are in series and each capacitor sees half of the applied potential, such that the peak-to-peak variation of mechanical energy in the capacitors becomes $U_{p-p,2,acc}$:

$$\begin{aligned} U_{p-p,2,acc}(V_{in}) &= 2\left(U_{acc}\left(-\frac{V_{in}}{2}\right) - U_{acc}\left(\frac{V_{in}}{2}\right)\right) \\ &= \frac{2\varepsilon_{SiN}^2 A}{t_{SiN}^3 E_{SiN}} \left(\frac{V_{fb} V_{in}^3}{4} + V_{fb}^3 V_{in}\right). \end{aligned} \quad (4)$$

We note that Eq. (3) and (4) yield null values if $V_{fb} = 0$, meaning that only harmonic actuation would be possible in this case.

In the depletion case (i.e., $\psi_m < \psi_s$, see Fig. 1d), three-dimensional charge distribution (see Fig. 1d) in the semiconductor imposes that we integrate the stress along the depletion region, which dimension $x_d$ is illustrated in Fig. 1(d) and given by [20]:

$$x_d(V_{in}) = -\frac{\varepsilon_{Si}}{C_{SiN}} + \sqrt{\left(\frac{\varepsilon_{Si}}{C_{SiN}}\right)^2 + \left(\frac{2\varepsilon_{Si}}{qN_A}\right)(V_{in} - V_{fb})}, \quad (5)$$

where $\varepsilon_{Si}$ is the dielectric constant of silicon, $q = 1.602 \times 10^{-19}$ C is the electron charge, and $N_A$ is the acceptor concentration in silicon. From continuum mechanics [17], the local stress $\sigma(x)$ at a given position $x$ in the junction is given by integrating electrostatic forces ($F(x)$) in the junction, starting from the stress-free position at $x = x_d$ (i.e., we assume no stress where the electrical field is null):

$$\sigma(x) = -\int_{x_d}^{x} dF(x), \quad (6)$$

with $dF(x) = -qN_A \cdot Y(x) \cdot dx$, where $Y(x)$ is the electric field and $-qN_A$ is the charge density. In the depletion region, $Y(x) = \frac{qN_A(x_d-x)}{\varepsilon_{Si}}$ [17], such that the integral yields:

$$\sigma(x) = \frac{(qN_A)^2}{\varepsilon_{Si}} \frac{(x_d - x)^2}{2}, for\ 0 < x < x_d. \quad (7)$$

By ensuring the continuity of stress (i.e., force balance) at the Si/SiN interface, and by assuming no stress outside of the junction region, we also find:

$$\sigma(x) = \begin{cases} \frac{(qN_A)^2}{\varepsilon_{Si}} \frac{x_d^2}{2}, for\ -t_{SiN} < x < 0. \\ 0, \quad for\ x < -t_{SiN}\ or\ for\ x > x_d. \end{cases} \quad (8)$$

The strain energy stored in a thickness $dx$ of the junction is given by $dU = \frac{1}{2}\sigma\epsilon\ d\forall$ with $d\forall = A \cdot dx$. By considering the stress ($\sigma$) – strain ($\epsilon$) relation for a Young's Modulus $E$, $\epsilon = \sigma/E$, we can integrate $dU = \frac{1}{2E} A\ \sigma(x)^2\ dx$ over the junction length (i.e., from $-t_{SiN}$ to $x_d$), to obtain the total strain energy stored in the junction, in the depletion regime ($U_{dep}$):

$$U_{dep}(x_d(V_{in})) = \frac{A(qN_A)^4(x_d)^4}{8(\varepsilon_{Si})^2}\left(\frac{t_{SiN}}{E_{SiN}} + \frac{x_d}{5E_{Si}}\right). \quad (9)$$

Note that this relation depends on the applied voltage $V_{in}$ through Eq. (5). In the depletion case, making the analysis for two capacitors in series is not as straightforward as for the accumulation case since the capacitor values are voltage-dependent; we therefore do not derive a simple expression as in Eq. (3) – (4) for the charge accumulation case.

Nevertheless, we can get a sense of which of the two regimes produces the most variation in strain energy by comparing Eq. (2) with Eq. (9), as shown in Fig. 2. We note that a capacitor in the accumulation regime stores more mechanical energy, which is likely due to the weaker junction capacitance occurring in the depletion regime.

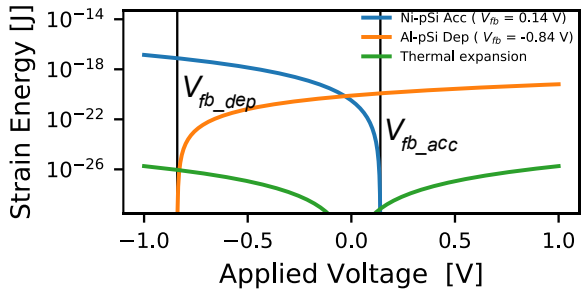

Figure 2. Theoretical stored strain energy vs. applied voltage for the charge accumulation model (Eq. 2), the depletion model (Eq. 9), and the thermal expansion thermal model (Eq. 15). A capacitor area of $2 \times 2$ mm$^2$ and a SiN layer thickness of 100 nm are used for the calculations. For the thermal model, an actuation frequency of 83 kHz is considered. Vertical lines indicate the flat band voltages ($V_{fb}$) in the depletion (dep) and accumulation (acc) regimes.

In a practical situation, such as for the experimental results presented below, the applied voltage ($V_{in}$) can typically extend beyond the flatband voltages of the capacitors. Capacitors can therefore undergo a change of regime, between accumulation and depletion, during each actuation cycle. It is therefore likely that the choice of metal (i.e., the flatband voltage) does not have a critical influence on actuation performances.

### III. THERMAL ANALYSIS

Current flowing in the silicon substrate may also cause an acoustic wave resulting from thermal expansion-induced mechanical stresses. The strain energy ($U_{th}$) due to a temperature change $\Delta T$ in a volume of silicon $\forall$ is given by:

$$U_{th} = \frac{1}{2}\forall E_{Si}\epsilon^2 = \frac{1}{2}\forall E_{Si}\alpha_{Si}^2\Delta T^2, \quad (10)$$

where $\alpha_{Si}$ is the thermal expansion coefficient of silicon and $\forall$ represents the effective silicon volume that expands under the effect of Joule heating.

Assuming a simple lumped thermal capacitance model [21], the amplitude of time-dependent temperature signal $\Delta T$ is proportional to the dissipated power $P_{in}$, and is filtered by a one-pole low-pass filter of characteristic response time $\tau_{th}$:

$$\Delta T = P_{in}R_{th}\frac{1}{1 + j\omega\tau_{th}}, \quad (11)$$

where $R_{th}$ is the thermal resistance, in units of [K/W], between the heated volume and its surrounding environment. The angular frequency of actuation is denoted by $\omega$, while the characteristic thermal response time is $\tau_{th} = R_{th}C_{th}$. In turn, the thermal capacitance $C_{th} = \forall\rho_{Si}c_{Si}$ (in units of J/K) is derived from the element volume $\forall$, its density ($\rho_{Si}$) and specific heat ($c_{Si}$). If we now assume that the thermal response is much slower than the actuation frequency (i.e., $\omega \gg 1/\tau_{th}$), Eq. (11) simplifies to a form a that does not depend on $R_{th}$:

$$\Delta T = \frac{P_{in}R_{th}}{j\omega R_{th}C_{th}} = \frac{P_{in}}{j\omega C_{th}}. \quad (12)$$

By substituting Eq. (12) into Eq. (10), we obtain the following expression of thermal strain as a function of the dissipated electrical power $P_{in}$:

$$U_{th} = \frac{1}{2}\forall E_{Si}\alpha_{Si}^2\left(\frac{P_{in}}{j\omega C_{th}}\right)^2. \quad (13)$$

Referring to the equivalent electrical circuit in Fig. 1 (b), we express the dissipated electrical power as $P_{in} = R_{Si}I^2$, where $R_{Si}$ is the electrical resistance of silicon between the contact pads, and $I$ is the current. We assume (with experimental evidence given later in Fig. 4), that the impedance of the two capacitors in series ($Z_c$) is the factor limiting the current $I$. Therefore, by using $I = \frac{V_{in}}{Z_c} = V_{in}\ j\omega\left(\frac{C_{SiN}A}{2}\right)$, the dissipated power $P_{in}$ can be rewritten as a function of the applied voltage:

$$P_{in} = R_{Si}I^2 = -R_{Si}V_{in}^2\omega^2\left(\frac{C_{SiN}A}{2}\right)^2, \quad (14)$$

where the minus sign originates from the $\pi$ phase shift between current and voltage signal in a capacitive impedance. Substituting in Eq. (13), the strain energy as a function of applied voltage signal is given by:

$$U_{th}(V_{in}) = \frac{1}{32} E_{Si} \alpha^2 \omega^2 V_{in}^4 \left(\frac{R_{Si}C_{SiN}^2 A^2}{C_{th}}\right)^2 \forall. \quad (15)$$

In Eq. (15), we can finally substitute geometrical parameters to estimate the magnitude of thermal expansion forces and compare them to the electrostatic forces discussed in the previous section. As a rough first-principle approximation, we approximate the interaction volume $\forall$ as the cylinder depicted in Fig. 1 (e), with $L \approx 4$ mm given by the spacing between the electrodes, and $A$ as the electrode area (e.g., 4 mm$^2$ in our typical configuration). We use $R_{Si} \approx \frac{\varrho_{el} \cdot L}{A}$ with $\varrho_{el}$ as the resistivity of silicon (4.5 $\Omega \cdot$ cm in the current work). Eq. (15) can then be rewritten as:

$$U_{th}(V_{in}) = \frac{1}{32} E_{Si} \alpha_{Si}^2 \omega^2 V_{in}^4 \left(\frac{\varrho_{el} \varepsilon_{SiN}^2}{t_{SiN}^2 \rho_{Si} c_{Si}}\right)^2 \forall. \quad (16)$$

Considering our interaction volume of ~16 mm$^3$ and an actuation frequency of 83 kHz (representative of our experimental conditions described below), values of $U_{th}$ are presented in Fig. 2 and are found to be very low compared with electrostatic strain. We therefore expect the contribution of thermal expansion forces to be negligible.

## IV. EXPERIMENTAL METHODS

The silicon nitride resonators characterized in this work are fabricated in-house using a custom process based on KOH etching. We employ commercially available p-type 100 mm diameter silicon wafers coated with a nominal 100 nm thickness of low stress LPCVD (low pressure chemical vapor deposition) SiN. The SiN thickness post-fabrication has been measured to ~90 nm by ellipsometry in [8].

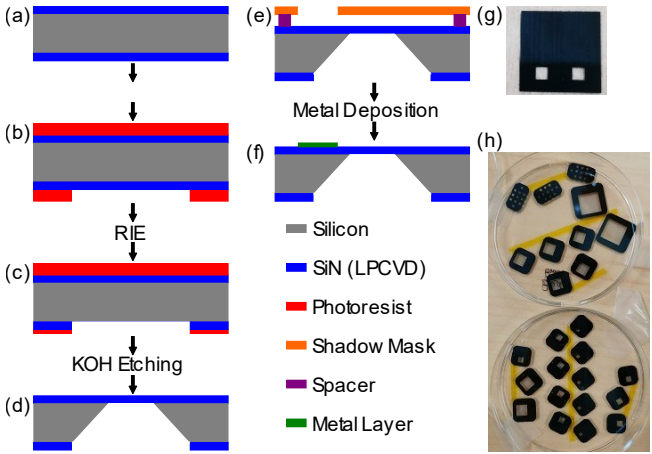

Figure 3. (a) 500 μm thickness (100) single crystal silicon wafer with low-stress LPCVD SiN coating. (b) Double-side photoresist (PR) spin coating and backside PR layer opening by photolithography. (c) Backside SiN film etching by reactive ion etching. (d) Hot KOH solution etches the silicon substrate from the backside to reveal the free-standing SiN membrane. (e) Alignment and attachment of a metal shim shadow mask. Spacers allow for a small distance that protects the membrane. (f) Metal deposition by electron beam evaporation to create the metal electrode pads. (g) 13.3 × 13.3 mm sheet metal shadow mask with 2 × 2 mm openings made by femtosecond laser machining. (h) Picture of various sizes of membrane chips fabricated in house from the same substrate.

Fig. 3 shows a general procedure for LPCVD SiN membrane chip fabrication. More details on the fabrication process are provided in supplementary information section S1. After protecting the wafer front side with photoresist, openings on the backside of the 100 nm wafer are defined using conventional photolithography and etched by reactive ion etching (RIE). KOH solution etches the silicon substrate from one side and reveals the SiN membrane on another side. This procedure allows the fabrication of membranes of various dimensions, ranging from $1 \times 1$ mm$^2$ up to $12 \times 12$ mm$^2$, that have been employed in some of our previous work [8,9]. In the present case, all characterized membranes are $1.7 \times 1.7$ mm.

Two types of MDS capacitors are experimentally investigated in the present work. The first type (Al-pSi, see Fig. 4b) is formed by depositing two $2 \times 2$ mm$^2$ aluminum pads on the silicon substrate using a shadow mask process and electron beam metal evaporation (see Fig. 3 e,f). The aluminum layer is 100 nm thick and comprises a 5 nm thick titanium adhesion layer. The shadow mask is fabricated by femtosecond laser etching of a 51 μm thick (2 mils) blue tempered high carbon steel shim stock by Precision Brand®. Details of the femtosecond laser etching process and the mask-chip mounting method are given in supplementary section S2. Electrical connection to the aluminum pads is achieved using 34-gauge silver coated copper wires (Accu-Glass Products Inc, part TYP23) attached to the aluminum pads using high vacuum-compatible Nickel Paste (PELCO® High Performance Nickel Paste and Product No. 16059, 16059-10). The second type of MDS capacitor (Ni-pSi, see Fig. 4c) is formed by attaching wires directly to the silicon nitride layer using the same nickel paste, but without the underlying aluminum contact pads.

MDS capacitors are excited by a lock-in amplifier providing a sinusoidal AC voltage signal centered on one of the membrane resonance eigenfrequencies. The excitation frequency is adjusted in real-time using a digital phase-locked-loop (PLL) that keeps the excitation signal centered on the membrane eigenfrequency. Membrane displacement is measured using an optical fiber interferometer forming a low finesse cavity between the optical fiber cleaved end and the membrane [22] (see Fig. 4a). The optical fiber is single mode at the laser wavelength (1563.52 nm), with a mode field diameter of ~10 μm. The experiment is conducted in high vacuum $(\sim 1 \times 10^{-6}$ hPa). Samples are mounted in the chamber using double-sided high vacuum compatible polyimide tape.

Prior to performing actuation experiments, we validate that our metal contacts are indeed capacitively coupled with the substrate. In Fig. 4 (d), we clearly see that the current flowing between the electrode scales with the drive frequency, which is a signature of a capacitive impedance. The magnitude of the current is much higher for the aluminum pads assembly. This is likely due to the much larger contact areas enabled by the deposited aluminum pads, resulting in a lower impedance. At high frequencies, the experimental curves deviate slightly from a purely linear behavior, which likely points out to a small resistive contribution of the substrate to the total impedance.

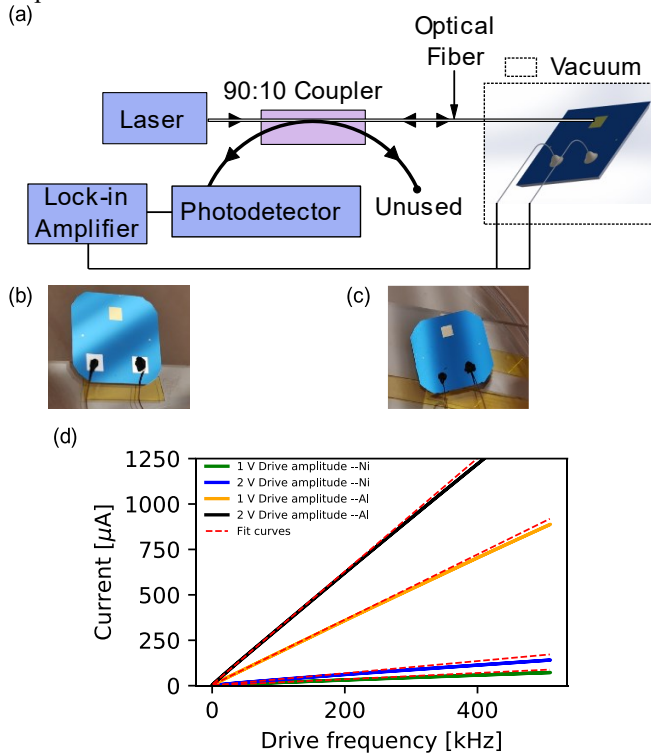

Figure 4. (a) Schematic illustration of the MDS capacitor actuation system driven by a lock-in amplifier and monitored with an optical fiber interferometer. (b, c) Metal electrodes are made of (b) evaporated aluminum pads for Al-pSi samples and (c) nickel paste applied directly on SiN for Ni-pSi samples. (d) Measured current flowing between electrodes vs. drive frequency for 1 V and 2 V drive voltages, and with nickel and aluminum pads.

## V. RESULTS AND DISCUSSION

Actuation results tend to confirm our theoretical prediction that charge accumulation (i.e., Ni-pSi in our case) is more effective than charge depletion (Al-pSi) for actuating SiN membranes. In Fig. 5, the Ni-pSi capacitor assembly achieves a 10 nm membrane actuation amplitude for a 2 V excitation signal, as opposed to 5 nm for the Al-pSi assembly. Although these amplitudes are on the same order of magnitude, the Ni-pSi achieves this result with a much smaller effective contact area (i.e., lower electrical power input), as indicated by the much lower current measurements in Fig. 4 (d).

We note that actuation amplitude decreases rapidly for higher frequency modes. This reduction can be partially explained by the expected increase in membrane stiffness ($k$) at high frequencies ($k = m_{eff}\omega^2$, where $m_{eff}$ is the membrane effective mass), or by changes of mechanical susceptibility resulting from variations of mechanical quality (Q) factors for the different modes (Q values are indicated in the legend in Fig. 5). Another possible explanation is the fact that our readout optical fiber is approximately positioned at the membrane center, which corresponds to displacement antinodes only for certain eigenmodes. Moreover, modes that nominally have an antinode at this position may also be hybridized to a different mode shape [23] and show smaller amplitudes.

A more thorough investigation of acoustic coupling of the electrodes with the substrate and the membrane would be necessary for predicting actuation amplitude as a function of frequency. We note that, at a typical frequency of ~200 kHz, the acoustic wavelength in silicon (~ 1.1 cm) is comparable to the chip dimension and the electrode spacing. Therefore, it is likely that these geometrical parameters play a significant role, which exact nature is beyond the scope of the present first-observation work. Likewise, the reason why some curves in Fig. 5 (c) present a non-linear response is not entirely clear. Possible effects include change of capacitance as a function of applied voltage or mode shape changes with the excitation voltage.

We find that the dissipated power in the chip is on the order of microwatts, which causes limited spurious chip heating. For example, considering the 80 kHz fundamental mode (1, 1) in the Ni-pSi case, the actuation power ($V_{in} \times I$) for a 2 V excitation is 56 µW if we use $I$ values from Fig. 4 (d). Moreover, due to the mostly capacitive impedance demonstrated in Fig. 4 (d), only a small fraction of this power is dissipated in the chip itself, while most of the power is dissipated in the external drive circuit. The power dissipated in the chip is given by:

$$P_{MDS} = R_{Si}I^2 \approx \varrho_{el}\frac{L}{A}I^2, \quad (17)$$

which is approximately 0.1 µW considering $L \approx 4$ mm, $A \approx 1$ mm$^2$, and the measured current from Fig. 4 d. In contrast, a typical shear piezoelectric stack (e.g., Thorlabs, PL5FBP3) routinely used for membrane excitation dissipate energy given as:

$$P \approx \frac{1}{8}\omega\,tan\delta\,C_{piezo}V_{in}^2, \quad (18)$$

where $tan\delta \approx 0.02$ is the dissipation factor, $f$ is the drive frequency, $C_{piezo} = 1.6\,nF$ is the piezoelectric chip capacitance. Considering the same membrane driving voltage and frequency, the dissipated power for a piezo actuator would be approximately 32 µW. Obviously, such comparative analysis is qualitative, since the voltage needed with a piezo actuator depends strongly on the interface between the piezoelectric actuator and a silicon chip, and is often in the mV or µV range. Nevertheless, the fact that the dissipated power values are on the same order of magnitude allows us to expect limited spurious heating from our capacitive actuation scheme. Spurious heating could also be minimized, if needed, by doping the silicon substrate to reduce resistive dissipation.

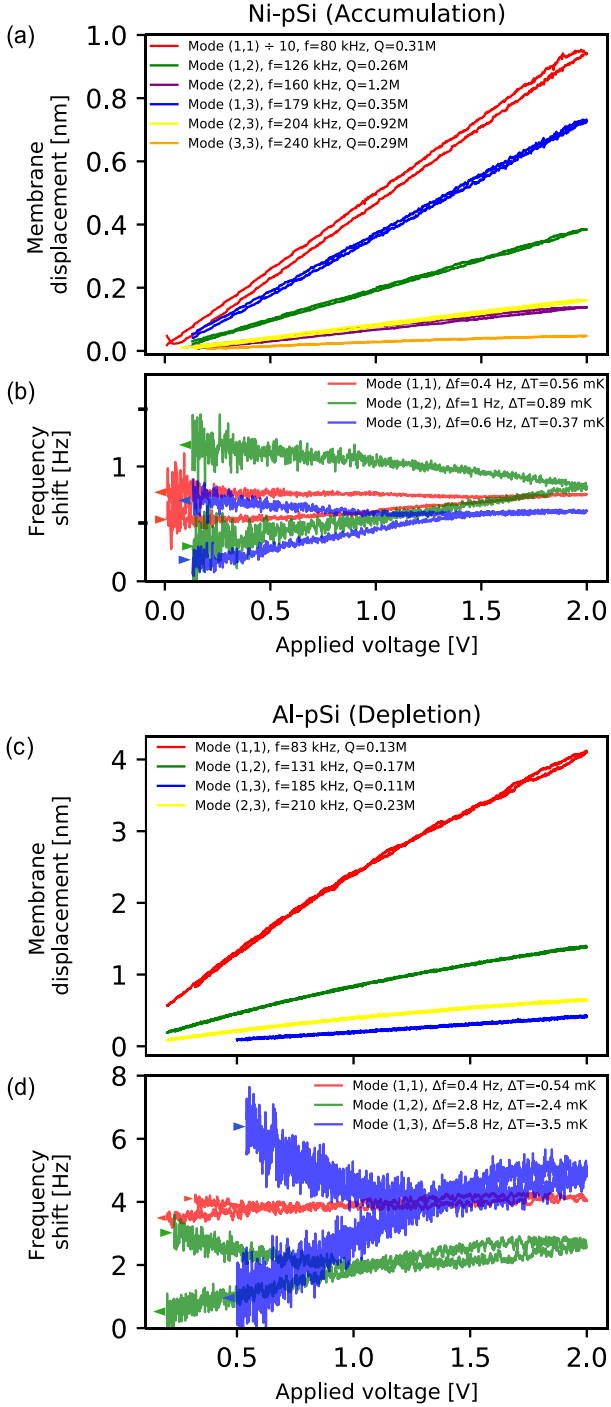

related to relative frequency shifts of the eigenmodes ($\Delta f/f_0$) using [24]:

$$\Delta T = \frac{2\Delta f(1-\nu)\sigma_{SiN}}{f_0 E_{Si}(\alpha_{Si} - \alpha_{SiN})} , \quad (19)$$

where $f_0$ is the original resonance frequency of the mode, $\sigma_{SiN} \approx 100$ MPa is the formation stress of SiN, $\nu$ is the Poisson ratio for SiN, and $\alpha$ denotes material thermal expansion coefficients. We note that all sweeps in this work present temperature variations of less than 4 mK, and that the most effective actuation assembly (Ni-pSi) results in less than 1 mK spurious heating. In Fig. 5 (d), the frequency shift indicates a temperature diminution during the sweep. This most likely results from a cool down from previous excitation at higher amplitude, for example, during adjustment of the PLL parameters. Typical duration each frequency sweep in Fig. 5 was 2 – 5 minutes, and spurious heating over longer actuation periods was not measured.

We finally observe that harmonic excitation can give better results than direct excitation in some cases. We define harmonic excitation by an excitation voltage centered at the mode eigenfrequency divided by an integer $n > 1$. Harmonic excitation results are shown in Fig. 6 (b), where harmonic 2 yields better performances than direct excitation for the Al-pSi assembly. Efficient harmonic excitation is not surprising given the polynomial dependence of the capacitor's energy with the applied voltage (see Eq. 2). A predictive model on the efficiency of harmonic excitation is, however, beyond the scope of this first-observation work.

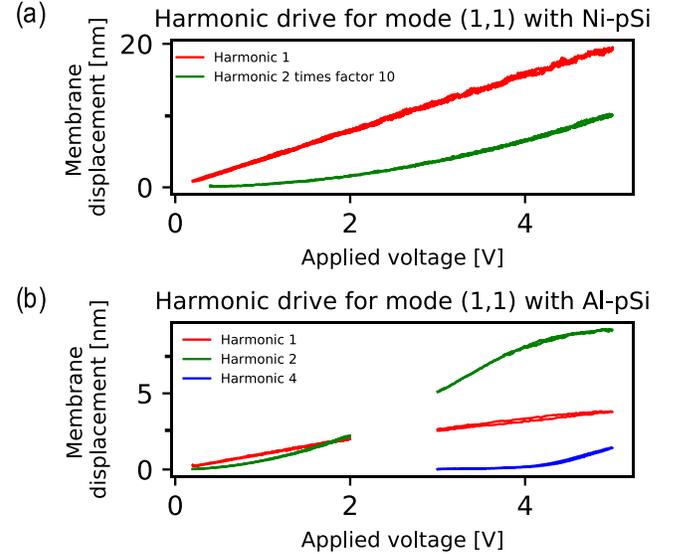

Figure 5: (a, c) Measured actuation amplitude as a function of actuation voltage, and (b, d) thermally-induced frequency-shifts for the Ni-pSi (top) and Al-pSi (bottom) assemblies. All sweeps are bi-directional with their direction indicated by arrows in (b, c). In panel (a) mode (1,1) amplitude is divided by 10 for better plot visibility.

To experimentally confirm limited chip heating during actuation, we measure (see Fig. 5 b, d) the eigenmode frequency shift occurring during the actuation sweeps of Fig. 5 (a, c). Temperature shifts of the substrate ($\Delta T$) can be

Figure 6: Harmonic actuation, whereas the excitation frequency is the membrane eigenmode frequency divided by the harmonic order. (a) Mode (1, 1) of the Ni-pSi assembly. (b) Mode (1, 1) for the Al-pSi assembly. In panel (a), the displacement amplitude for harmonic 2 is multiplied by a factor of 10 for better visibility. In panel (b), the measurements were conducted in two different sweeps (0.2-2 V and 3-5 V) which explain the discontinuities.

## VI. CONCLUSION

We have observed the actuation of silicon nitride nanomechanical resonators by exploiting electrostatic attraction force in metal-dielectric-capacitors formed in the underlying silicon frame. We have also developed first principle analytical models, which are in qualitative agreement with our experimental observations. As predicted by our models, the charge accumulation regime (Ni-pSi) is more efficient at actuating the membrane resonators compared to charge depletion (Al-pSi).

The greatest advantage of our method is most likely its remarkable simplicity. Applying metal paste on the chip frame and connecting it with wires suffices to build the actuation device, providing that the substrate is a dielectric-coated semiconductor. In future work, multiple paths exist for optimizing our method by investigating the effects of electrodes, membrane, and silicon frame geometries. Placing the electrodes closer to the membrane could potentially limit acoustic propagation loss, leading to stronger actuation. Also, smaller chips could potentially confine the acoustic excitation better and allow for stronger actuation. Acoustic interference effects inside the silicon frame would also need to be accounted for and are likely to play a significant role, especially since the acoustic wavelength in silicon is comparable to the chip dimensions at several frequencies of interest. Finally, reducing the distance between the electrodes or increasing silicon doping could potentially decrease the substrate electrical resistance, reducing power dissipation and spurious heating.

**REFERENCE:**

[1]  V. P. Adiga, B. Ilic, R. A. Barton, I. Wilson-Rae, H. G. Craighead, and J. M. Parpia, *Approaching Intrinsic Performance in Ultra-Thin Silicon Nitride Drum Resonators*, Journal of Applied Physics **112**, 064323 (2012).

[2]  D. J. Wilson, C. A. Regal, S. B. Papp, and H. J. Kimble, *Cavity Optomechanics with Stoichiometric SiN Films*, Phys. Rev. Lett. **103**, 207204 (2009).

[3]  V. P. Adiga, B. Ilic, R. A. Barton, I. Wilson-Rae, H. G. Craighead, and J. M. Parpia, *Modal Dependence of Dissipation in Silicon Nitride Drum Resonators*, Appl. Phys. Lett. **99**, 253103 (2011).

[4]  B. Nair, A. Naesby, and A. Dantan, *Optomechanical Characterization of Silicon Nitride Membrane Arrays*, Opt. Lett. **42**, 1341 (2017).

[5]  J. Park, H. Qin, M. Scalf, R. T. Hilger, M. S. Westphall, L. M. Smith, and R. H. Blick, *A Mechanical Nanomembrane Detector for Time-of-Flight Mass Spectrometry*, Nano Lett. **11**, 3681 (2011).

[6]  P. Peng, A. S. Sezen, R. Rajamani, and A. G. Erdman, *Novel MEMS Stiffness Sensor for Force and Elasticity Measurements*, Sensors and Actuators A: Physical **158**, 10 (2010).

[7]  N. Hartgenbusch, M. Borysov, R. Jedermann, and W. Lang, *Characterization and Design Evaluation of Membrane-Based Calorimetric MEMS Sensors for Two-Dimensional Flow Measurement*, IEEE Sensors J. **20**, 4602 (2020).

[8]  N. Snell, C. Zhang, G. Mu, A. Bouchard, and R. St-Gelais, *Heat Transport in Silicon Nitride Drum Resonators and Its Influence on Thermal Fluctuation-induced Frequency Noise*, ArXiv:2110.00080 [physics.app-ph], (2021).

[9]  M. Giroux, C. Zhang, N. Snell, G. Mu, M. Stephan, and R. St-Gelais, *High Resolution Measurement of Near-Field Radiative Heat Transfer Enabled by Nanomechanical Resonators*, Appl. Phys. Lett. **119**, 173104 (2021).

[10]  V. P. Adiga, R. De Alba, I. R. Storch, P. A. Yu, B. Ilic, R. A. Barton, S. Lee, J. Hone, P. L. McEuen, J. M. Parpia, and H. G. Craighead, *Simultaneous Electrical and Optical Readout of Graphene-Coated High Q Silicon Nitride Resonators*, Appl. Phys. Lett. **103**, 143103 (2013).

[11]  M. Kurek, M. Carnoy, P. E. Larsen, L. H. Nielsen, O. Hansen, T. Rades, S. Schmid, and A. Boisen, *Nanomechanical Infrared Spectroscopy with Vibrating Filters for Pharmaceutical Analysis*, Angew. Chem. Int. Ed. **56**, 3901 (2017).

[12]  K. Y. Fong, W. H. P. Pernice, M. Li, and H. X. Tang, *High Q Optomechanical Resonators in Silicon Nitride Nanophotonic Circuits*, Appl. Phys. Lett. **97**, 073112 (2010).

[13]  Q. P. Unterreithmeier, E. M. Weig, and J. P. Kotthaus, *Universal Transduction Scheme for Nanomechanical Systems Based on Dielectric Forces*, Nature **458**, 1001 (2009).

[14]  X. Zhou, S. Venkatachalam, R. Zhou, H. Xu, A. Pokharel, A. Fefferman, M. Zaknoune, and E. Collin, *High- Q Silicon Nitride Drum Resonators Strongly Coupled to Gates*, Nano Lett. acs.nanolett.1c01477 (2021).

[15]  Y. Wu, G. Ding, C. Zhang, J. Wang, S. Mao, and H. Wang, *Design and Implementation of a Bistable Microcantilever Actuator for Magnetostatic Latching Relay*, Microelectronics Journal **41**, 325 (2010).

[16]  B. R. Ilic, S. Krylov, M. Kondratovich, and H. G. Craighead, *Optically Actuated Nanoelectromechanical Oscillators*, IEEE J. Select. Topics Quantum Electron. **13**, 392 (2007).

[17]  T. Tsushima, Y. Asahi, H. Tanigawa, T. Furutsuka, and K. Suzuki, *Various Vibration Modes in a Silicon Ring Resonator Driven by p–n Diode Actuators Formed in the Lateral Direction*, Jpn. J. Appl. Phys. **57**, 067201 (2018).

[18]  J. H. T. Ransley, A. Aziz, C. Durkan, and A. A. Seshia, *Silicon Depletion Layer Actuators*, Appl. Phys. Lett. **92**, 184103 (2008).

[19]  S. W. P. van Sterkenburg, *The Electrostriction of Silicon and Diamond*, J. Phys. D: Appl. Phys. **25**, 996 (1992).

[20] C. Hu, *Modern Semiconductor Devices for Integrated Circuits* (Prentice Hall, Upper Saddle River, NJ., 2010).
[21] T. Bergman, A. Lavine, F. Incropera. *Fundamentals of heat and mass transfer*, (John Wiley & Sons, Inc., Hoboken, NJ., 2019) 8th ed.
[22] D. Rugar, H. J. Mamin, and P. Guethner, *Improved Fiber-optic Interferometer for Atomic Force Microscopy*, Appl. Phys. Lett. **55**, 2588 (1989).
[23] S. Chakram, Y. S. Patil, L. Chang, and M. Vengalattore, *Dissipation in Ultrahigh Quality Factor SiN Membrane Resonators*, Phys. Rev. Lett. **112**, 127201 (2014).
[24] M. Wang, R. Zhang, R. Ilic, V. Aksyuk, and Y. Liu, *Frequency Stabilization of Nanomechanical Resonators Using Thermally Invariant Strain Engineering*, Nano Lett. **20**, 3050 (2020).
[25] R. Tahvildari, PhD, University of Ottawa (2017).

# Observation of Silicon Nitride Nanomechanical Resonator Actuation Using Capacitive Substrate Excitation

## (Supplementary Information)

Gengyang Mu[1,2], Nikaya Snell[1,2], Chang Zhang[1,2], Xitong Xie[1,2], Radin Tahvildari[3,4], Arnaud Weck[1,2,3], Michel Godin[3,4], Raphael St-Gelais[2,2,3*]

[1]Department of Mechanical Engineering, University of Ottawa, Ottawa, ON, Canada
[2]Centre for Research in Photonics, University of Ottawa, Ottawa, ON, Canada
[3]Department of Physics, University of Ottawa, Ottawa, ON, Canada
[4]Ottawa-Carleton Institute for Biomedical Engineering, Ottawa, ON, Canada

**S1: SiN Membrane Fabrication**

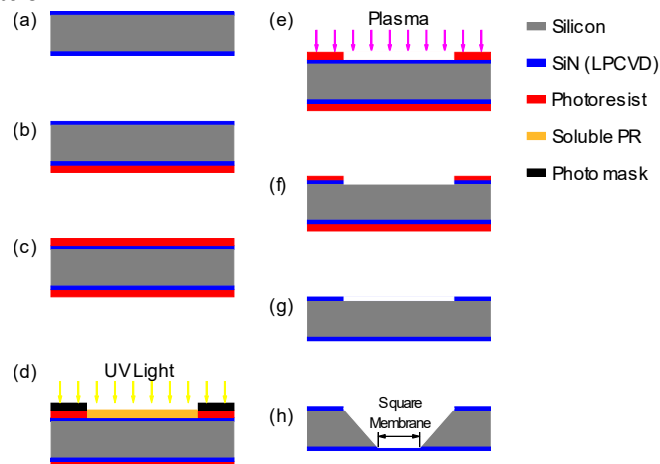

Figure S1: (a) Original wafer coated with 100 nm low stress LPCVD SiN. (b) Front side PR coating on SiN membrane side for protection. (c) Backside PR coating for further etching. (d) Expose PR partially with photomask under UV light. (e) RIE plasma etching. (f) Wafer after RIE with SiN film opening partially. (g) Remove rest of PR. (h) Wafer after the hot KOH etching with free-standing SiN membranes.

Fig. S1 shows a general procedure for LPCVD SiN membrane chip fabrication starting from a four-inch (100) single-crystal 500 μm thick silicon wafer coated on both sides with 100 nm LPCVD SiN (see Fig. S1 a). The wafer is first spin coated with S1811 PR on both sides (Fig. S1b, c) with 2000 rpm spinning speed for achieving approximately 1.3 μm PR thickness. The backside PR coating is used for photolithographic processing, while the frontside coating is used solely to protect the SiN device layer from scratches and contaminations during the fabrication. Both PR coating processes use 1 min, 115 °C soft bake heating process to polymerize the PR. A photomask is then used, in hard contact lithography mode, to expose the PR layer and duplicate the photomask pattern (Fig. S1. d). After exposure, the exposed PR area becomes soluble in the developer—i.e., MF-321, as instructed in the S1800 PR series spec sheet. During the developing process, the wafer is immersed in the developer for 1 min to remove soluble PR, followed by a one-bath DI water rinse. The wafer is then etched (see Fig. S1e) by reactive ion etching (RIE) with etching conditions of 25 sccm $CF_4$, 5 sccm $O_2$, 100 W input power, 8 Pa pressure, in a SAMCO's RIE-10NR™ reactor. After 2 mins of RIE etching, the 100 nm SiN film is entirely removed in exposed areas (see Fig. S1f). The remaining PR is removed on both sides of the wafer, after RIE, by ~30 seconds immersion in acetone (Fig. S1g). An isopropyl alcohol (IPA) rinse is used to remove acetone's residue, after which the samples are ready for KOH etching (Fig. S1h). During the whole procedure, soft tip tweezers in Fig. S2h are used in order to reduce the chance of physical damages (e.g., scratches, impacts, shocks etc.) done to SiN films and photoresist (PR) layers.

*raphael.stgelais@uottawa.ca

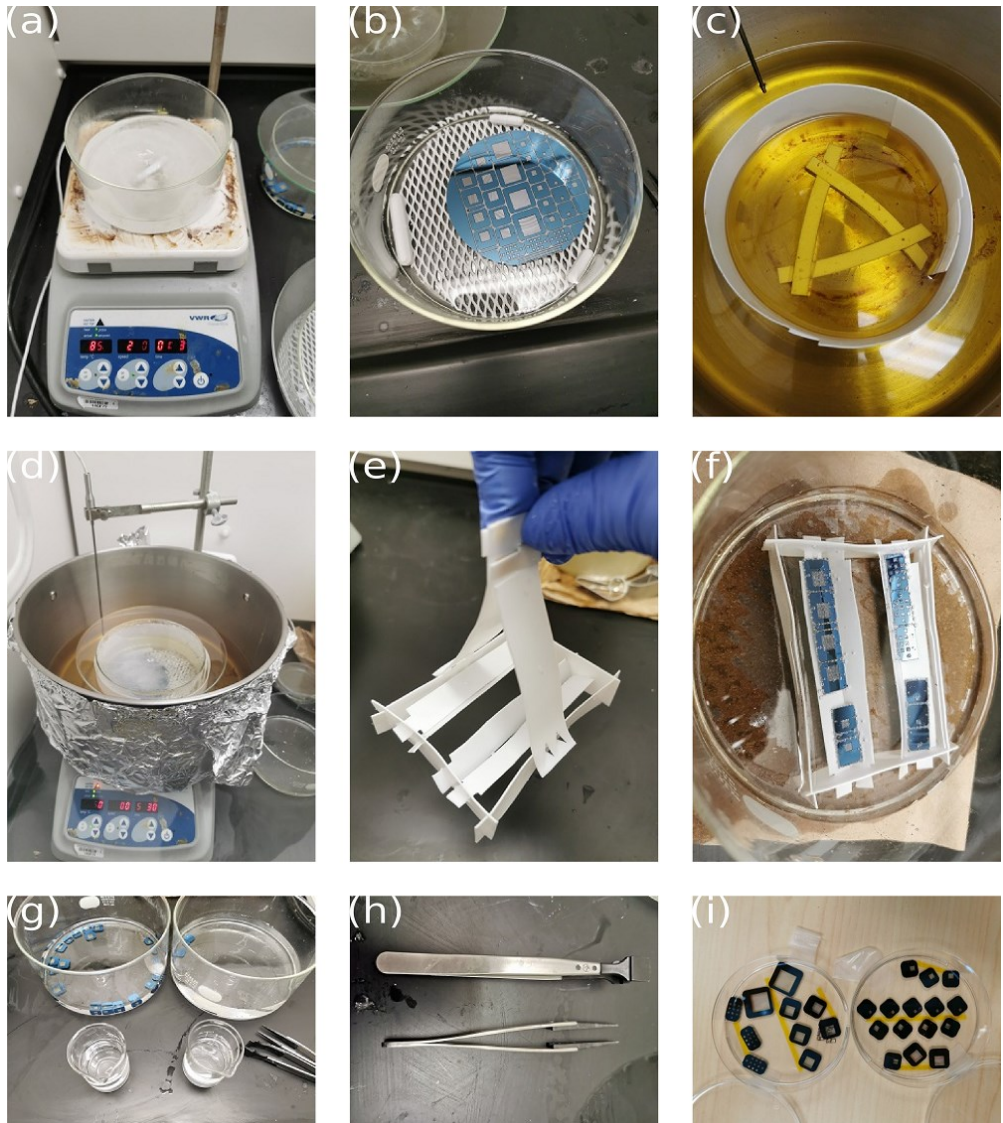

Figure S2: (a) KOH solution mixing. (b) Set up for the first half hot KOH etching process with the whole wafer. (c) The Teflon spacer and baker holder. (d) Oil bath set up. (e) The Teflon membrane chip rack and rack lifter. (f) Set up for second half hot KOH etching process with membrane chip strips. (g) Rinse set up with DI water, acetone and IPA. (h) Soft tip tweezers. (i) Membrane chips from the same substrate in sample boxes.

Stable etching results in KOH are ensured by immersion heating of the etching solution in an oil bath [25]. Immersion heating allows for a more even distribution of heat in the solution, as opposed to direct heating of KOH solution on a hot plate that results in a large temperature gradient in the solution. An oil bath is preferred over a water bath due to its higher boiling point. Corn oil is selected because of its wide availability and its good stability at high temperatures. Fig. S2 shows the oil bath for hot KOH etching. From separate calibrations with an immersion thermometer at 60 – 85 °C oil temperatures, we find that the oil bath is typically 5 °C warmer than the KOH solution, in a steady state. The oil temperature is therefore monitored throughout the process to be at a temperature of 5 °C higher than the desired KOH solution temperature. The KOH etch solution is not directly monitored and is covered by a concave-shaped glass lid (Fig. S2d) that recirculates evaporated water back to the solution, keeping KOH concentration steady during etching.

The KOH solution is prepared in a glass beaker (PYREX® 150 × 75, No.3140) with 120 g pure KOH salt and 280 ml room temperature 18.2 MΩ-cm DI water, for a 30% (w/w) mixture. The mixing process is assisted by magnetic stirring with 200 rpm stirring speed (see Fig. S2a). The hot KOH solution will eventually attack the glass beaker, but at a relatively low etching rate. As a safety precaution, we therefore replace the beaker after every five KOH etch batches (i.e., five completed wafers). Since KOH dissolution in water is exothermic, the solution temperature rises to ∼50 °C during the mixing process. The oil bath is therefore pre-heated to a similar temperature, such that we prevent thermal stress when the two are put in contact. Spacers (see

Fig. S2c) made of polytetrafluoroethylene (PTFE) are used for avoiding direct heating of the KOH solution by the hot plate. A thin PTFE sleeve (see Fig. S2c) also prevents random beaker movement in the oil bath.

Since SiN Membranes are facing downwards during etching, a PTFE mesh at the bottom of the KOH beaker (see Fig. S2b) protects the membranes from direct contact with the bottom of the beaker. The mesh also provides spacing between the beaker and the wafer for the manipulation of chips with tweezers. The etching reaction generates $H_2$ bubbles which will stick to the mesh and can make it float out of the solution. To prevent this, three PTFE weights are placed on the mesh surrounding the wafer to keep the wafer and mesh immersed. The membrane faces downwards in order to let $H_2$ bubbles float up, preventing interruption of the etching process seized by bubbles trapped under the wafer.

Hot KOH etching is completed in two steps. The first step is conducted on the entire wafer, and its goal is to etch the wafer approximately halfway through. This takes 2 - 2.5 hours at 80 °C etching temperature (i.e., 85 °C oil bath temperature). The wafer is then separated into strips (See Fig. S2f) by cleaving in ~550 μm wide grooves defined by the KOH etch. In the second step, a custom-made PTFE etch rack is used to hold separated chip strips, as shown in Fig. S2f. The rack provides easier access to the chips, and stabilizes them at the end of the process when the chips become individually separated by the KOH etch. A PTFE rack lifter shown in Fig. S2e helps move the whole rack with membrane chips in and out of the solution. The second etching step lasts about 1.5 to 2 hours at 80 °C etching temperature. In some cases, such as for large area $12 \times 12$ mm$^2$ membranes, the etching temperature is decreased to 60°C (i.e., 65 °C oil bath temperature) for slower, more delicate, etching. In this particular case, the etch time for the second step is 4 – 5 hours.

After hot KOH etching, membrane chips are rinsed twice in DI water, once in acetone, and once in IPA (Fig. S2h). The double DI water rinse in large size glass beakers maximizes the removal of KOH residue and SiN film fragments. In this step, the PTFE rack, with multiple membrane chips on it, is lifted out of the KOH solution and immersed into DI water entirely with the rack lifter. During acetone and IPA rinses (in smaller beakers for minimizing solution evaporation), chips are manipulated individually. Finally, chips are blown dry with an $N_2$ gas gun, out of the IPA solution, and then stored.

Using this process, membranes survival rates are relatively high, as indicated by Table S1 for our most recent membrane fabrication run. Survival rates depend a lot on operator experience—the vast majority of membrane destruction occurs during operator manipulation. This indicates that the process itself does not induce significant damages to the membrane. For the particular case of Table S1, the operator had prior experience with seven prior KOH etchings, and all membrane destructions occurred during sample manipulation in the rinsing steps. To reduce the chances of damages, we find that soft tip tweezers are better suited for manipulation. They also reduce Si dust that forms during handling, compared to regular steel tweezers.

Table S1: Latest membrane chip fabrication survival rates on different sizes of membranes.

| Sample type (square side length) | Sample quantity (expectation) | Survival rate |
|---|---|---|
| 1 mm | 3 | 100% |
| 1.5 mm | 5 | 100% |
| 3 mm | 8 | 100% |
| 6 mm | 9 | 77.8% |
| 12 mm | 4 | 50% |

In order to secure membranes for transportation between labs, chips are mounted onto double-sided tape at one edge with membrane side facing upward in plastic sample boxes (See Fig. S2i).

**S2: Shadow Mask**

The shadow masks used in this work are fabricated using the direct laser writing (etching) method. The material is blue tempered high carbon steel shim stock (Precision Brand®), with a thickness of 51 μm. The material is first cleaned using cleanroom tissue to reduce debris during machining, and to improve the uniformity of the machining results. To machine the steel surface, a Yb:KGW femtosecond laser source (PHAROS from light Conversion, Inc.) was used, which operates at a laser wavelength of 1030 nm. All the shadow masks are machined with the same laser parameters. The pulse duration, total laser power, and repetition rate are fixed at 300 fs, 360 mW, and 10 kHz, resulting in pulse energy of 36 μJ. The machining laser was focused on the steel surface through a telecentric f-theta lens (Sill Optics) to obtain a laser spot size of about 20 microns. The laser beam path is fixed during the machining, and the substrate (steel shim) is mounted on a stage (Aerotech) to enable X-Y-Z movement. The machining speed at which the substrate moves relative to the laser is 0.1 mm/s, resulting in a total laser fluence of $1.8 \times 10^4$ J / cm$^2$. The machining procedure is repeated twice for each shadow mask, to improve the smoothness of the machined edges. This is proven to be helpful in reducing the chances of damaging the SiN surfaces.

Since laser cutting leaves a residue, the shadow mask is first cleaned with cleanroom wipes to prevent contamination of the membranes. It then goes through ultrasonic cleaning in DI water at 80 kHz, 30% of total power, for 5 min with Elmasonic P 30 H Ultrasonic cleaning unit.

## S3: Metal deposition

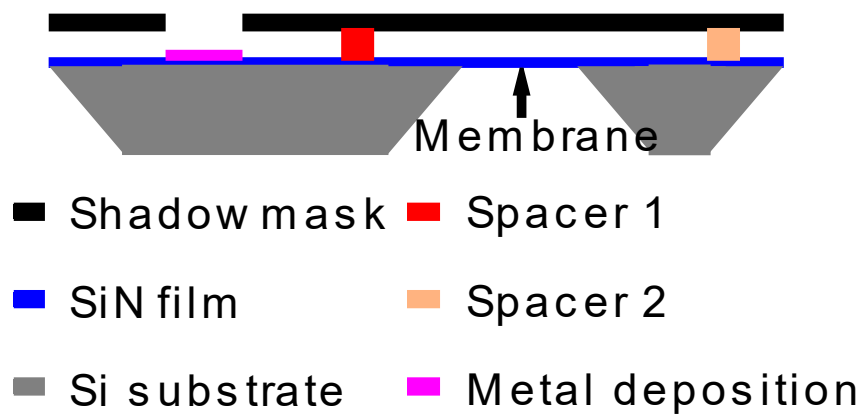

Figure S3: Shadow mask mounting for metal deposition.

The shadow mask is mounted by a small piece of double-sided tape (i.e., spacer 1) on the membrane chip. A piece of inactivated double-sided tape (i.e., double-sided tape with an unpeeled cover as Spacer 2) prevents contact between the mask and the membrane. This spacer is inactivated to facilitate shadow mask removal at the end of the process. Then, the membrane chip sample goes through 5 nm titanium (Ti) and 100 nm aluminum (Al) metal deposition in an electron beam metal evaporator, with 0.1 nm/s and 0.2 nm/s deposition rate respectively. Ti acts as an adhesion layer between Al and SiN.

Once the metal deposition is completed, the mask and double-sided tapes are removed, and 34-gauge silver coated copper wires (Accu Glass Products Inc, part TYP23) is attached to the Al electrode pads using high vacuum-compatible Nickel Paste (PELCO® High Performance Nickel Paste and Product No. 16059, 16059-10), for the Al-pSi depletion case. The accumulation case (Ni-pSi) only required electrical wire attachment with nickel paste directly on SiN, without the prior shadow-mask and metal evaporation step. Polyimide insulation is removed from the copper wires' tip by manual ablation with sandpaper. Hot plate heating of the chip at 50 °C is used during wire attachment such that the paste solidifies almost instantly upon contact with the chip. On the hot plate, chips are mounted on a heavier glass slide by double-sided tape to prevent spurious movement during wire attachment. Uncured nickel paste is taken from its container using a toothpick and is then placed on the wire tip. In order to help the uncured paste stick with the wires, the wire tip is bent into a hook shape. The coated wire tip is then manually put in contact with the chip at 50 °C. The bonds solidify partially in a few seconds, and another clean toothpick is used to lightly press down the paste onto the chip to help with this initial solidification. After this process, the paste undergoes the manufacturer recommended curing sequence for achieving final electrical and mechanical properties: 2 – 4 hours curing at room temperature followed by 2 hours curing at 93 °C. During these curing periods, an aluminum foil cover is used to shield chips and avoid contamination.